\pdfoutput=1
\documentclass{PoS}

\usepackage[utf8]{inputenc}
\usepackage[numbers]{natbib}
\usepackage[multidot]{grffile}
\usepackage{fixltx2e}

\usepackage[kerning=true,protrusion=true,expansion=true]{microtype}
\usepackage[caption=false]{subfig}
\usepackage{amsmath}

\usepackage{units}
\usepackage{bbm}

\DeclareMathAlphabet{\mathcal}{OMS}{cmsy}{m}{n}

\global\long\def\One{\mathbbm{1}}

\global\long\def\spec{\operatorname{spec}}
\global\long\def\sign{\operatorname{sign}}

\title{Staggered domain wall fermions}

\ShortTitle{Staggered domain wall fermions}

\author{\speaker{Christian Hoelbling}\\
        Department of Physics, University of Wuppertal, D-42119 Wuppertal, Germany\\
        E-mail: \email{hch@uni-wuppertal.de}}
        
\author{Christian Zielinski\\
        Division of Mathematical Sciences, Nanyang Technological University, Singapore 637371 \&
        Department of Physics, University of Wuppertal, D-42119 Wuppertal, Germany\\
        E-mail: \email{zielinski@pmail.ntu.edu.sg}}
        
\abstract{We construct domain wall fermions with a staggered kernel and investigate
their spectral and chiral properties numerically in the Schwinger
model. In some relevant cases we see an improvement of chirality by
more than an order of magnitude as compared to usual domain wall fermions.
Moreover, we present first results for four-dimensional quantum chromodynamics,
where we also observe significant reductions of chiral symmetry violations
for staggered domain wall fermions.}

\FullConference{34th annual International Symposium on Lattice Field Theory\\
		24-30 July 2016\\
		University of Southampton, UK}

\begin{document}

\section{Introduction}

When implementing chiral fermions on the lattice, domain wall fermions
\cite{Kaplan:1992bt,Shamir:1993zy,Furman:1994ky} are a well-known
alternative to the computationally very expensive overlap construction
\cite{Ginsparg:1981bj,Hasenfratz:1998ri,Luscher:1998pqa}. By means
of massive interacting fermions in $d+1$ dimensions, one can formulate
lattice fermions with an approximate chiral symmetry in $d$ dimensions.
The accuracy of this approximate symmetry is controlled by the extent
of the extra dimension, where the limit of an infinite extent can
be concisely described by an overlap operator. Domain wall fermions
are computationally cheaper, permitting a simpler approach to parallelization
and allowing for an easier tunneling between different topological sectors
as compared to overlap fermions. This, however, comes at the price of
replacing the exact chiral symmetry of overlap fermions with an approximate
one.

Traditionally, domain wall fermions are used with a Wilson kernel as
the naïve application of the formalism to the staggered kernel
fails due to the lack of some technical properties within the staggered
framework. This problem was eventually overcome with Adams' proposal
of staggered domain wall fermions \cite{Adams:2009eb,Adams:2010gx},
where one introduces a modified kernel operator, i.e.\ so-called
staggered Wilson fermions. The resulting novel kernel operator is constructed
by adding a suitable flavored mass term \cite{Golterman:1984cy} to
staggered fermions (see also Ref.\ \cite{deForcrand:2012bm}).
As staggered Wilson fermions appear to be computationally
more efficient than the Wilson kernel \cite{Adams:2013tya}, one can
hope for a computationally cheaper and possibly more chiral formulation.
Besides staggered domain wall fermions, staggered Wilson fermions
also permit the formulation of staggered overlap fermions \cite{Adams:2011xf,deForcrand:2011ak}
with a well-defined index \cite{Adams:2013lpa}.

In this report, we present selected results of our investigations of
the spectral properties and chiral symmetry violations of
domain wall fermions with Wilson and staggered Wilson kernel operators
in the Schwinger model \cite{Schwinger:1962tp}
as recently discussed in Ref.\ \cite{Hoelbling:2016qfv}. In addition,
we present first results for the case of four-dimensional quenched
quantum chromodynamics.

\section{Formulation}

We begin with a quick review of the $d$-dimensional kernel operators
and the $\left(d+1\right)$-dimensional domain wall fermion construction
($d=2,4$). We denote the lattice spacing by $a$ in the first $d$
dimensions and by $a_{d+1}$ in the extra dimension. The $\gamma_{\mu}$
matrices ($\mu=1,\dots,d$) refer to a representation of the Dirac
algebra, where we use the notation $\gamma_{d+1}$ for the chirality
matrix.

\paragraph*{Kernel operators.}

We denote the $d$-dimensional Wilson Dirac operator with bare fermion
mass $m_{\mathsf{f}}$ by $D_{\mathsf{w}}\left(m_{\mathsf{f}}\right)$.
For the definition of staggered Wilson fermions, we use our notation
introduced in Ref.\ \cite{Hoelbling:2016qfv} and write the Dirac
operator as $D_{\mathsf{sw}}\left(m_{\mathsf{f}}\right)=D_{\mathsf{st}}+m_{\mathsf{f}}+W_{\mathsf{st}}$.
Here $D_{\mathsf{st}}=\eta_{\mu}\nabla_{\mu}$ is the usual staggered
Dirac operator and our choice of the staggered Wilson term can be
compactly written as 
\begin{equation}
W_{\mathsf{st}}=\frac{r}{a}\left(\One+\lambda W_{\mathsf{st}}^{1\cdots d}\right),\qquad\lambda=\left(-1\right)^{\frac{d+2}{2}}.
\end{equation}
In four dimensions our $W_{\mathsf{st}}$ equals Adams' staggered
Wilson term, for the two-dimensional case we refer to our discussion
in Ref.\ \cite{Hoelbling:2016qfv}. In the following discussion,
we set the $d$-dimensional lattice spacing to $a=1$ and the (staggered)
Wilson parameter to $r=1$.

\paragraph*{Bulk operators.}

The bulk Dirac operator in the standard (i.e.\ original) formulation
of domain wall fermions reads
\begin{equation}
\overline{\Psi}D_{\mathsf{dw}}\Psi=\sum_{s=1}^{N_{s}}\overline{\Psi}_{s}\left[D_{\mathsf{w}}^{+}\Psi_{s}-P_{-}\Psi_{s+1}-P_{+}\Psi_{s-1}\right],
\end{equation}
where the extent of the extra dimension is denoted by $N_{s}$, the
$\left(d+1\right)$-dimensional fermion fields $\overline{\Psi}$ and
$\Psi$ have an index $s$ for the additional spatial coordinate,
we introduced $D_{\mathsf{w}}^{\pm}=a_{d+1}D_{\mathsf{w}}\left(-M_{0}\right)\pm\One$,
the domain wall height is given by $M_{0}$ and the chiral projectors
are defined as $P_{\pm}=\left(\One\pm\gamma_{d+1}\right)/2$. In the
extra dimension we impose the boundary conditions
\begin{equation}
P_{+}\left(\Psi_{0}+m\Psi_{N_{s}}\right)=0,\qquad P_{-}\left(\Psi_{N_{s}+1}+m\Psi_{1}\right)=0,
\end{equation}
where the fermion mass is controlled by the parameter $m$. Finally,
by letting $q=P_{+}\Psi_{N_{s}}+P_{-}\Psi_{1}$ and $\overline{q}=\overline{\Psi}_{1}P_{+}+\overline{\Psi}_{N_{s}}P_{-}$,
we can define $d$-dimensional fermion fields from the boundary.

Besides this original formulation, we also consider two common $\mathcal{O}\left(a_{d+1}\right)$
modifications. Boriçi's construction \cite{Borici:1999zw} follows
from the replacement 
\begin{equation}
P_{+}\Psi_{s-1}\to-D_{\mathsf{w}}^{-}P_{+}\Psi_{s-1},\qquad P_{-}\Psi_{s+1}\to-D_{\mathsf{w}}^{-}P_{-}\Psi_{s+1},
\end{equation}
while Chiu introduces additional weight factors for the formulation
of optimal domain wall fermions \cite{Chiu:2002ir} in order to improve
chiral properties.

\paragraph*{Effective operators.}

\begin{figure}[t]
\begin{centering}
\subfloat[Wilson kernel]{\includegraphics[width=0.45\textwidth]{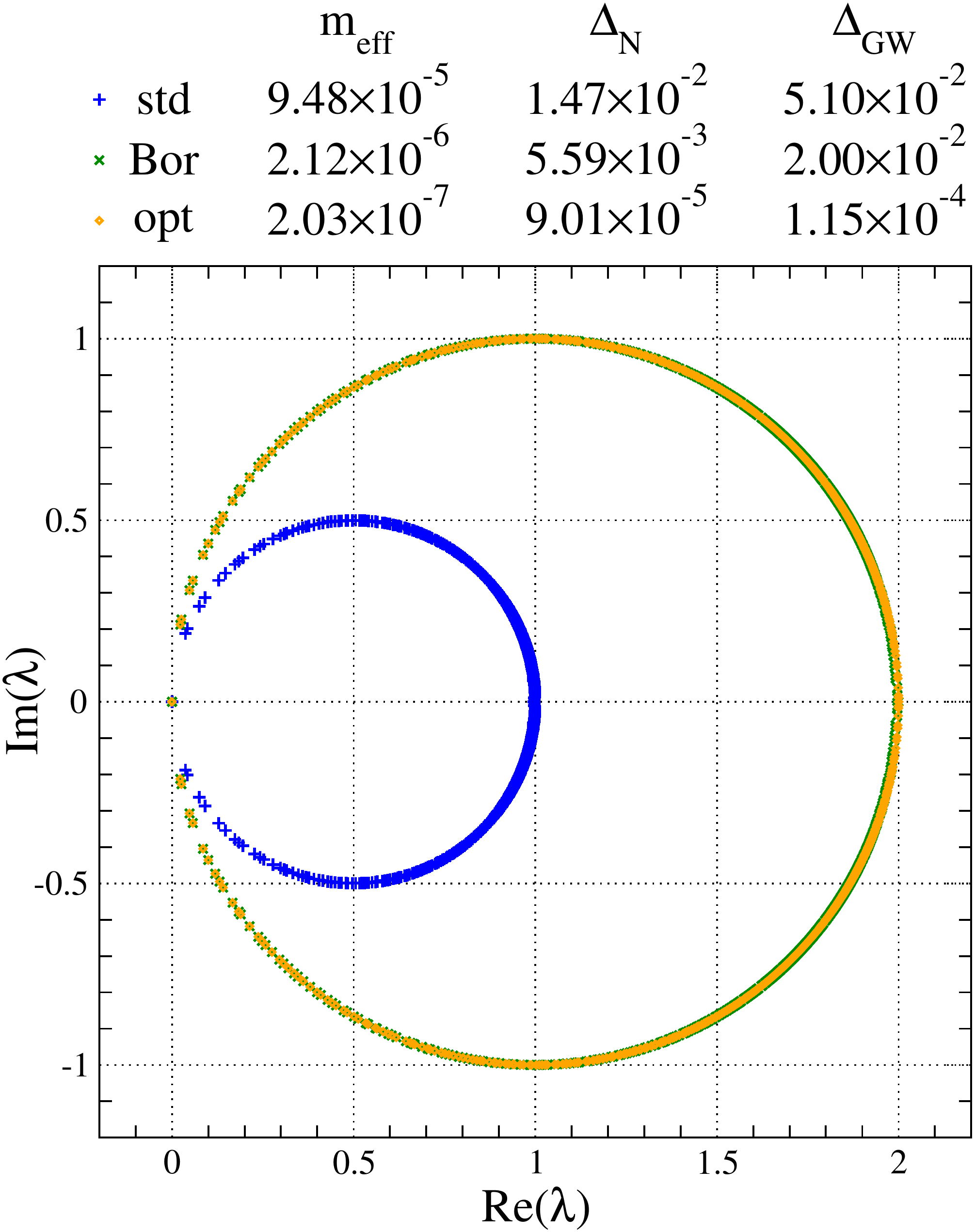}

}\hfill{}\subfloat[Staggered Wilson kernel]{\includegraphics[width=0.45\textwidth]{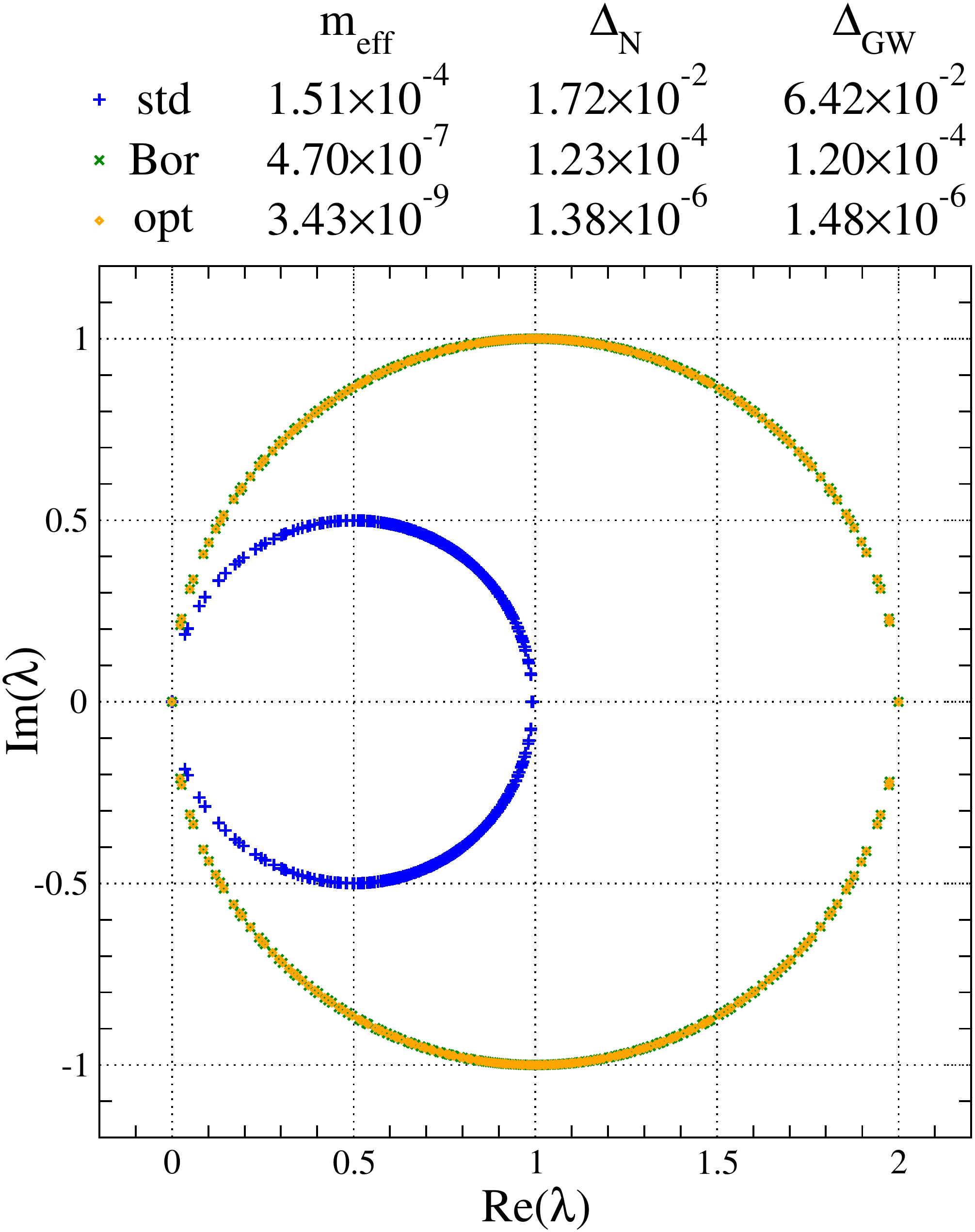}

}
\par\end{centering}
\caption{Spectrum of $\varrho D_{\mathsf{eff}}$ for the standard (std), Boriçi
(Bor) and optimal (opt) construction at $N_{s}=8$ on a $20^{4}$
configuration with topological charge $Q=3$ in the Schwinger model at $\beta=5$. \label{fig:gauge-deff-Ns8}}
\end{figure}

Following the derivation given in Refs.~\cite{Kikukawa:1999sy,Borici:1999zw}
and integrating out $N_{s}-1$ heavy modes, we can define a $d$-dimensional
low-energy effective action $S_{\mathsf{eff}}=\sum_{x}\overline{q}\left(x\right)D_{\mathsf{eff}}\,q\left(x\right)$,
where $D_{\mathsf{eff}}^{-1}\left(x,y\right)=\left\langle q\left(x\right)\overline{q}\left(y\right)\right\rangle $.
Note that in order to cancel the diverging contributions from the
heavy fermions in the chiral limit $N_{s}\to\infty$, one introduces
suitable pseudofermion fields. The effective operator can then be
written in closed form as
\begin{equation}
D_{\mathsf{eff}}=\frac{1+m}{2}\One+\frac{1-m}{2}\gamma_{d+1}\frac{T_{+}^{N_{s}}-T_{-}^{N_{s}}}{T_{+}^{N_{s}}+T_{-}^{N_{s}}},\label{eq:DefDeff}
\end{equation}
where $T_{\pm}=\One\pm a_{d+1}H$. In the case of the standard construction
$H$ equals the modified kernel operator $H_{\mathsf{m}}=\gamma_{d+1}D_{\mathsf{m}}\left(-M_{0}\right)$
with $D_{\mathsf{m}}=D_{\mathsf{w}}/\left(2\cdot\One+a_{d+1}D_{\mathsf{w}}\right)$.
For Boriçi's construction, we find the standard kernel $H_{\mathsf{w}}=\gamma_{d+1}D_{\mathsf{w}}\left(-M_{0}\right)$
and that the effective Dirac operator equals the polar decomposition
approximation of Neuberger's overlap operator. Finally, for Chiu's
optimal construction the fraction on the right hand side of Eq.\ \eqref{eq:DefDeff}
is replaced by Zolotarev's optimal rational function approximation
\cite{zolotarev1877application} of $\sign H_{\mathsf{w}}$. In all
cases the corresponding overlap operator is given by $D_{\mathsf{ov}}=\lim_{N_{s}\to\infty}D_{\mathsf{eff}}$.

As discussed in full detail in Ref.\ \cite{Hoelbling:2016qfv}, in
order to ensure the same scale of all effective operators, we let
$\varrho=2\omega$ with
\begin{equation}
\omega=\begin{cases}
M_{0}-\frac{1}{2}a_{d+1}M_{0}^{2} & \textrm{for the standard construction},\\
M_{0} & \textrm{for Boriçi's/Chiu's construction}
\end{cases}
\end{equation}
and consider $\varrho D_{\mathsf{eff}}$ and $\varrho D_{\mathsf{ov}}$
in all our numerical investigations.

\paragraph*{Staggered versions.}

As shown in Ref.\ \cite{Adams:2010gx}, one can obtain a staggered
version of domain wall fermions by means of a replacement rule. In
our setting, it takes the form $D_{\mathsf{w}}\to D_{\mathsf{sw}}$,
$\gamma_{d+1}\to\epsilon$ with $\epsilon\left(x\right)=\left(-1\right)^{x_{1}/a+\dots+x_{d}/a}$,
allowing a full generalization of the previous discussion to the staggered
case.

\section{Numerical results}

For the numerical part of our work, we begin by introducing measures
of chirality followed by the discussion of selected results in the setting of the
Schwinger model and four-dimensional quantum chromodynamics. For the
remainder of this report, we consider the massless case $m=0$.

\paragraph*{Measures of chirality.}

In order to quantify chiral symmetry violations for the effective
Dirac operators, we define three different measures. The first is
the effective mass $m_{\mathsf{eff}}$ as used in Ref.\ \cite{Gadiyak:2000kz}.
Using periodic boundary conditions and considering a topologically
nontrivial configuration, we let
\begin{equation}
m_{\mathsf{eff}}=\min_{\lambda\in\spec H}\left|\lambda\right|=\min_{\Lambda\in\spec D^{\dagger}D}\sqrt{\Lambda}.
\end{equation}
As chiral properties imply the normality of the Dirac operator \cite{Kerler:1999dk},
we also consider deviations from operator normality as given by $\Delta_{\mathsf{N}}=\left\Vert \left[D,D^{\dagger}\right]\right\Vert _{\infty}$.
Finally, we can directly measure violations of the Ginsparg-Wilson
relation by
\begin{equation}
\Delta_{\mathsf{GW}}=\left\Vert \left\{ \gamma_{d+1},D\right\} -\omega^{-1}D\gamma_{d+1}D\right\Vert _{\infty},
\end{equation}
where in the staggered case $\gamma_{d+1}$ is replaced by $\epsilon$.
We note that for the overlap operator, all our measures vanish in exact
arithmetics.

\paragraph*{Schwinger model.}

\begin{figure}[t]
\begin{centering}
\hfill{}\subfloat[Without smearing]{\includegraphics[width=0.4\textwidth]{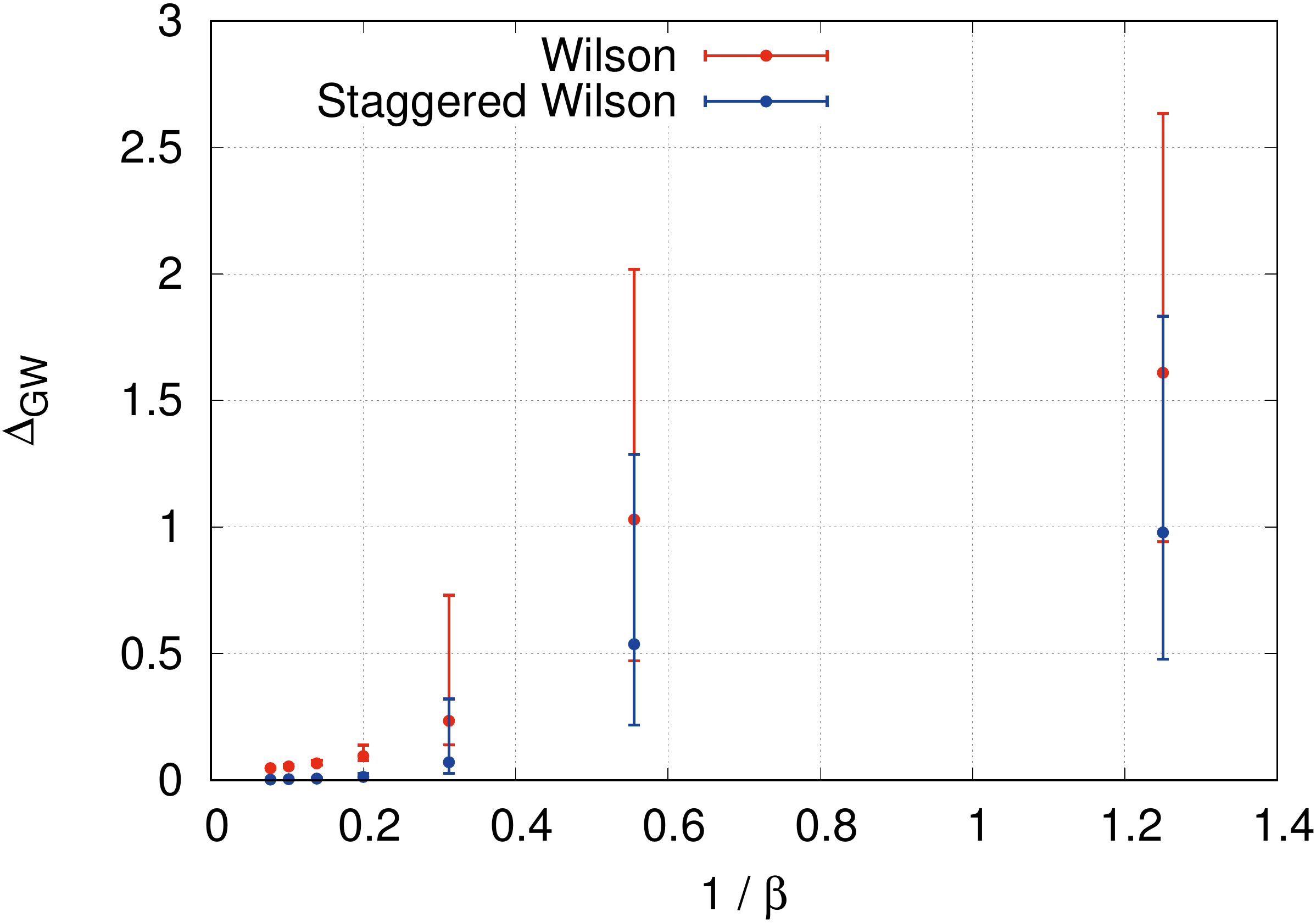}

}\hfill{}\subfloat[With smearing]{\includegraphics[width=0.4\textwidth]{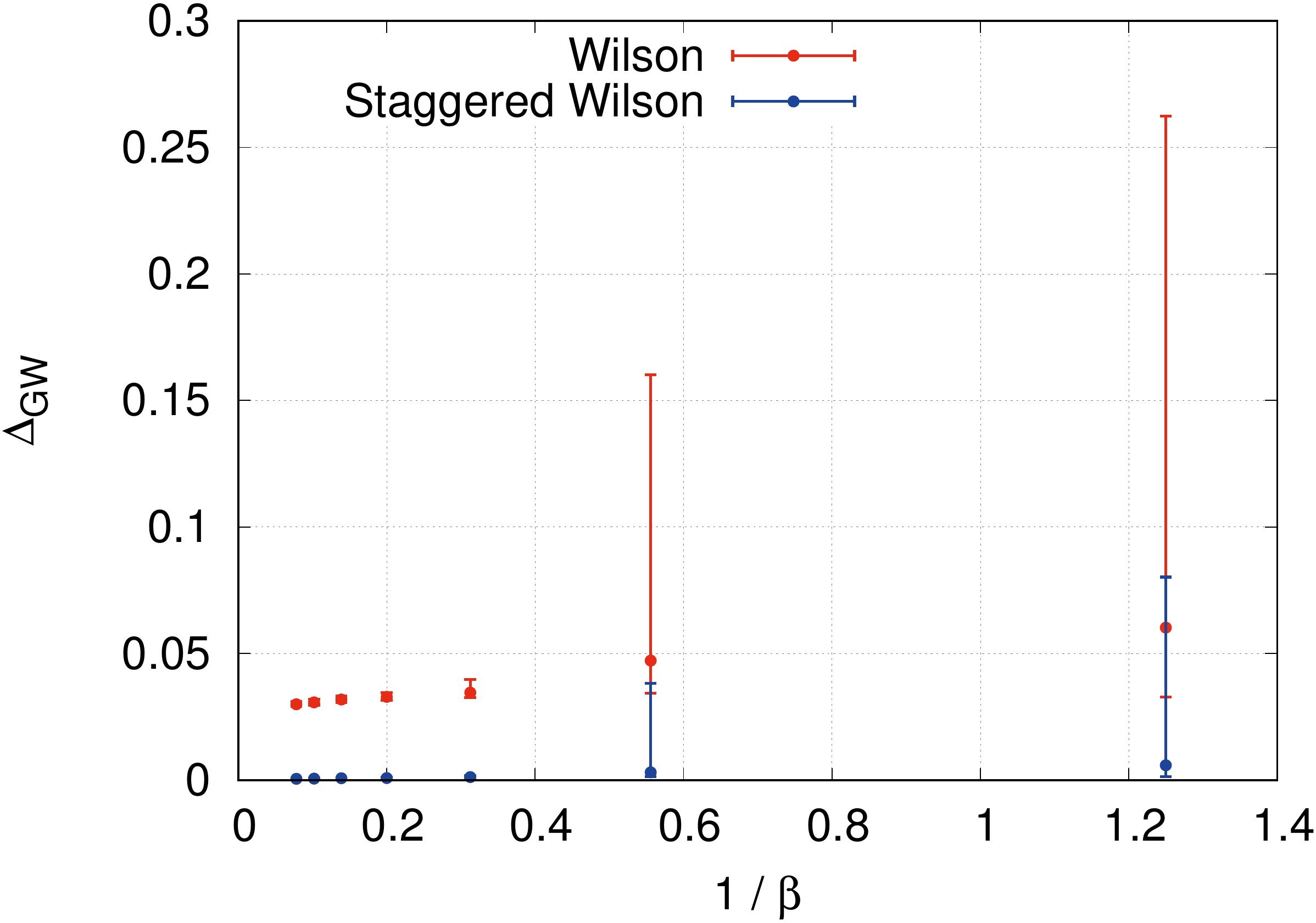}

}\hfill{}
\par\end{centering}
\caption{Measure $\Delta_{\mathsf{GW}}$ for $\varrho D_{\mathsf{eff}}$ in
the optimal construction at $N_{s}=4$ in the Schwinger model. \label{fig:cont-limit-gwr}}
\end{figure}
\begin{figure}[t]
\begin{centering}
\hfill{}\subfloat[Without smearing]{\includegraphics[width=0.4\textwidth]{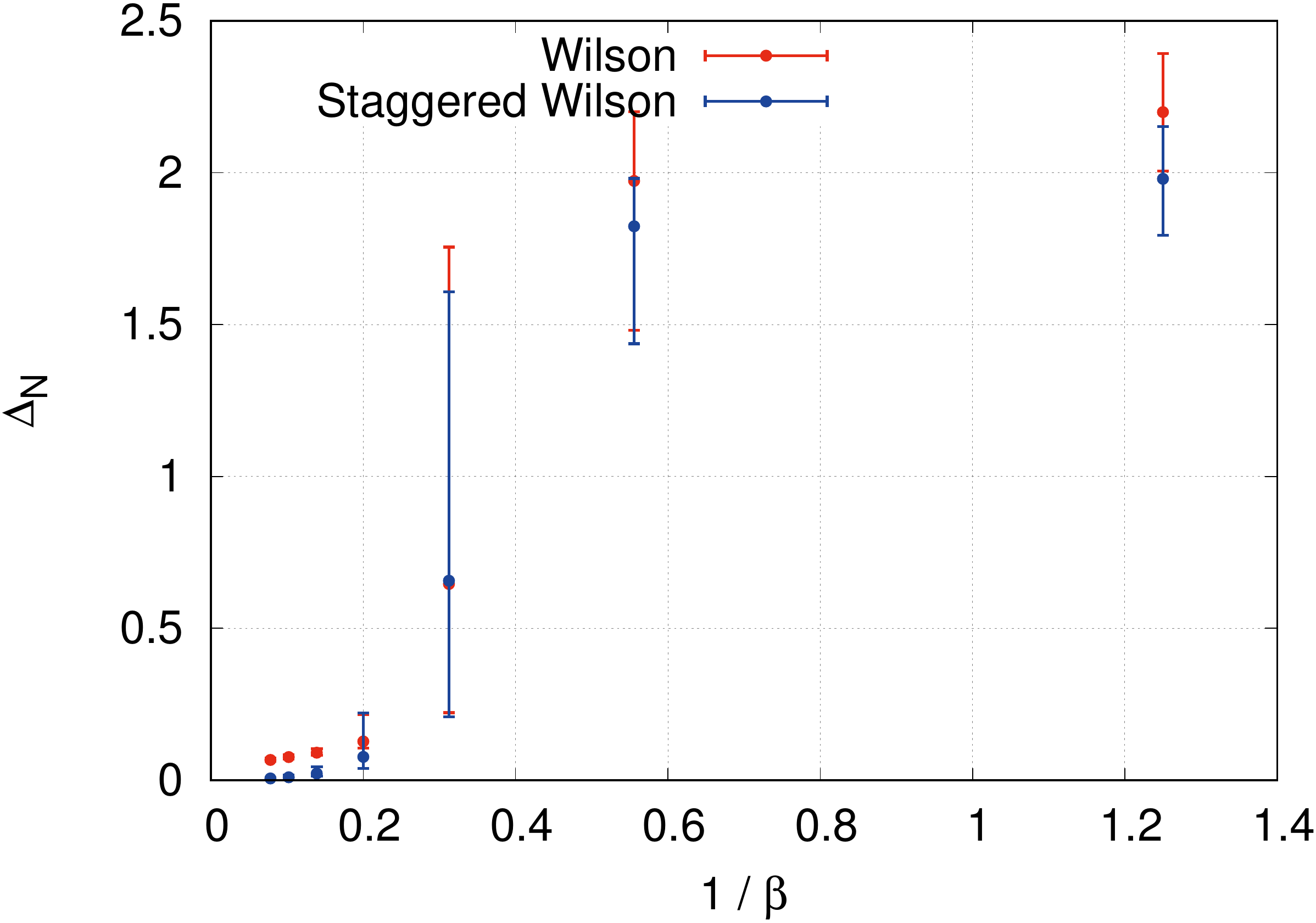}

}\hfill{}\subfloat[With smearing]{\includegraphics[width=0.4\textwidth]{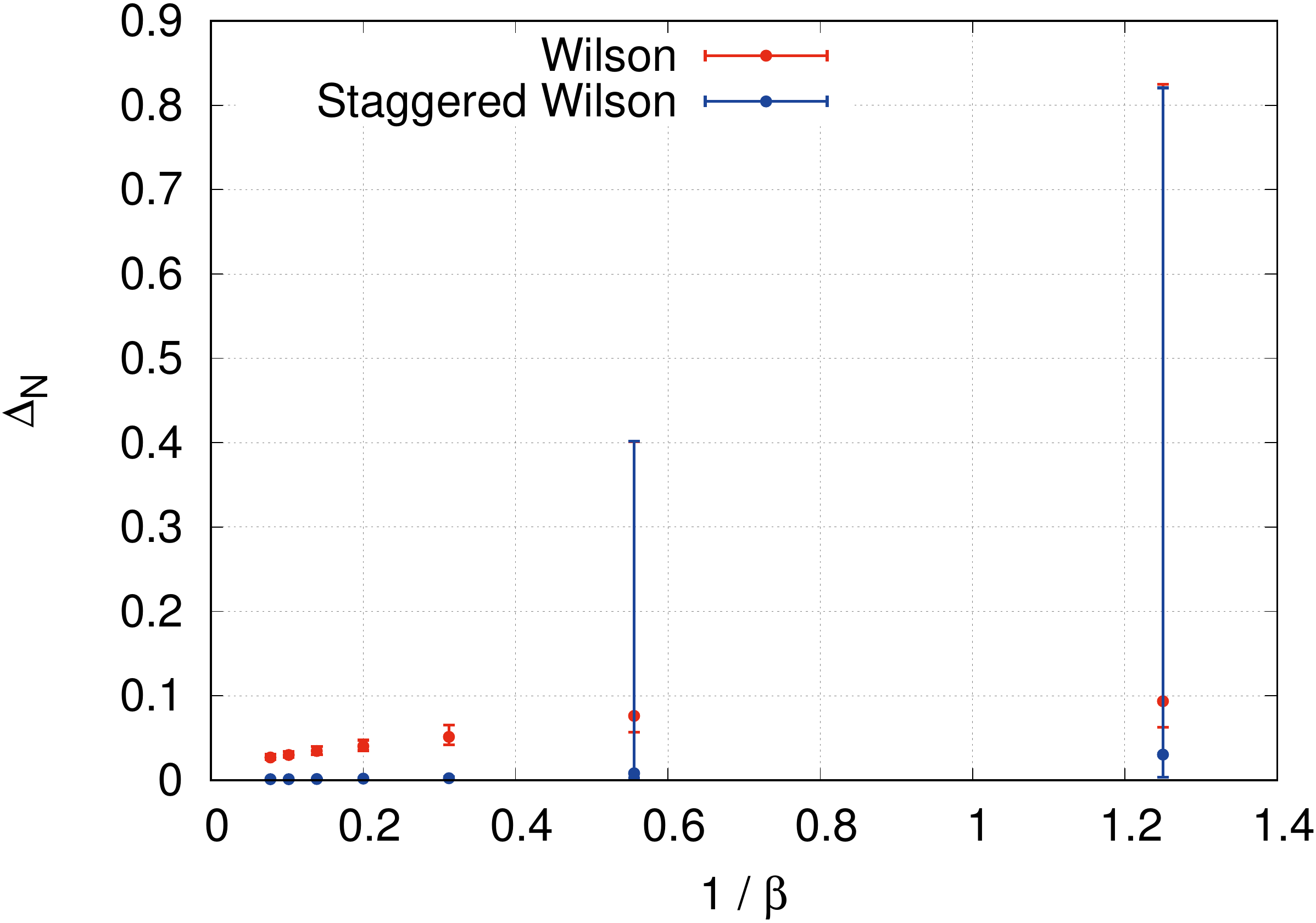}

}\hfill{}
\par\end{centering}
\caption{Measure $\Delta_{\mathsf{N}}$ for $\varrho D_{\mathsf{eff}}$ in
Boriçi's construction at $N_{s}=4$ in the Schwinger model. \label{fig:cont-limit-norm}}
\end{figure}

Let us now discuss a few selected results for the two-dimensional
Schwinger model. For our numerical study we use the canonical choice
$M_{0}=1$.

In Fig.\ \ref{fig:gauge-deff-Ns8}, we find the eigenvalue spectra
of the effective operator $\varrho D_{\mathsf{eff}}$ for various
constructions on an exemplary gauge configuration. We observe that already
for the small value of $N_{s}=8$ the spectrum very closely resembles
that of the corresponding overlap operator. Comparing the effective
operators with the usual Wilson and the staggered Wilson operator,
we note that in the case of the standard construction $m_{\mathsf{eff}}$,
$\Delta_{\mathsf{N}}$ and $\Delta_{\mathsf{GW}}$ are typically of comparable
magnitude. For Boriçi's and Chiu's optimal construction,
however, we find that the staggered construction shows notably improved
chiral properties compared to the Wilson case.

We are particularly interested in the chiral properties when approaching
the continuum limit. To this end we generated seven ensembles with
the setup of Ref.\ \cite{Durr:2003xs} with $10^{3}$ configurations
each, namely $8^{2}$ at $\beta=0.8$, $12^{2}$ at $\beta=1.8$,
$16^{2}$ at $\beta=3.2$, $20^{2}$ at $\beta=5.0$, $24^{2}$ at
$\beta=7.2$, $28^{2}$ at $\beta=9.8$ and $32^{2}$ at $\beta=12.8$.
The values for $\beta$ were chosen so that the physical volume is
kept fixed. We consider both unsmeared and three-step \textsc{ape}
smeared configurations \cite{Falcioni:1984ei} with a smearing parameter
of $\alpha=0.5$, resulting in $14\,000$ configurations in total.

In Figs.\ \ref{fig:cont-limit-gwr} and \ref{fig:cont-limit-norm},
we find two examples for the behavior of $\Delta_{\mathsf{GW}}$ 
and $\Delta_{\mathsf{N}}$ when $\beta$ is varied. In the figures, we
plot the median value and the $\unit[68.3]{\%}$-width of the distribution.
We find that domain wall fermions with a staggered Wilson kernel show
clearly superior chiral properties in the Schwinger model in the limit
of large $\beta$, where in some cases chiral symmetry violations
are reduced by more than an order of magnitude.

\paragraph*{Quenched quantum chromodynamics.}

\begin{figure}[t]
\begin{centering}
\subfloat[Wilson kernel]{\includegraphics[width=0.45\textwidth]{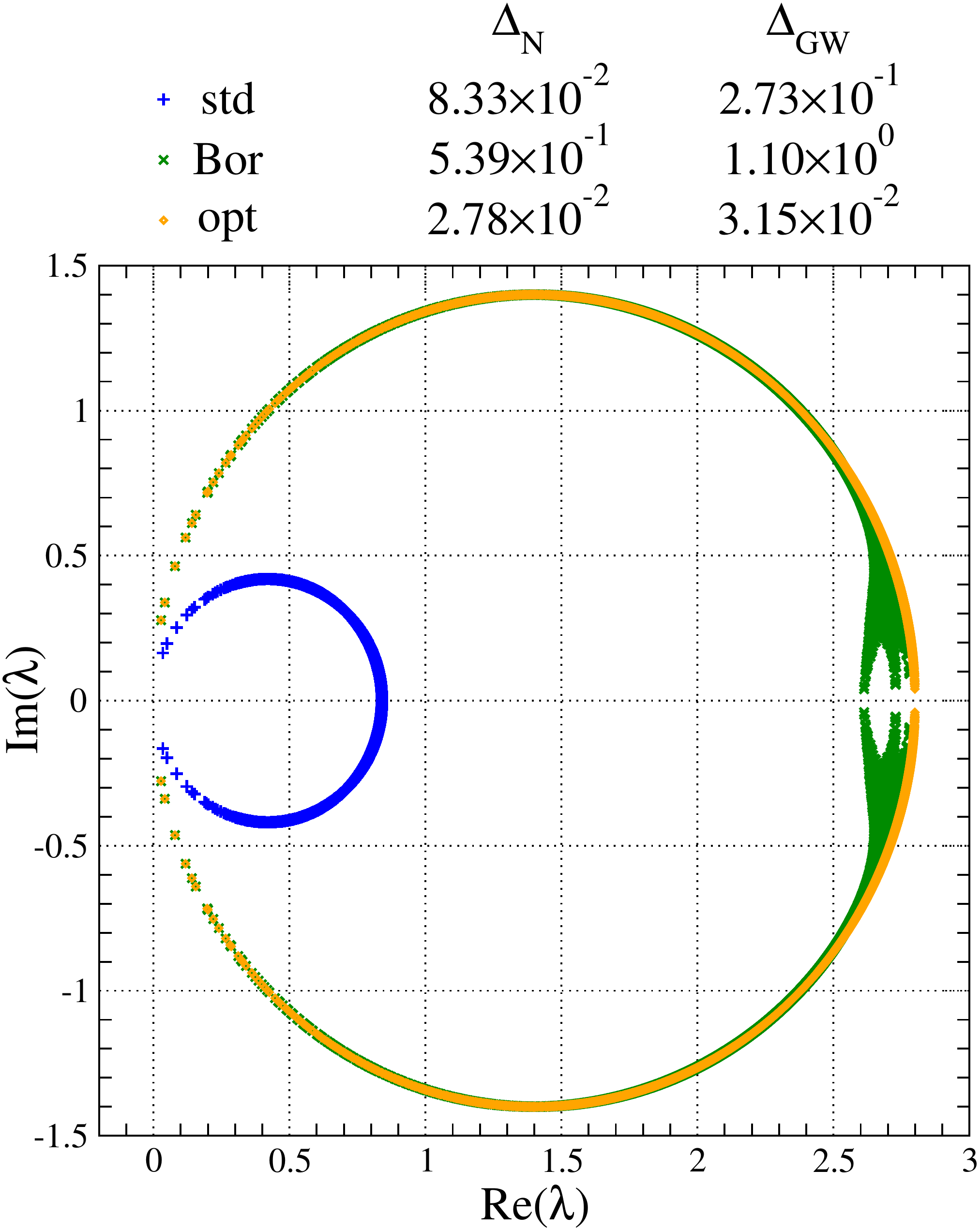}

}\hfill{}\subfloat[Staggered Wilson kernel]{\includegraphics[width=0.45\textwidth]{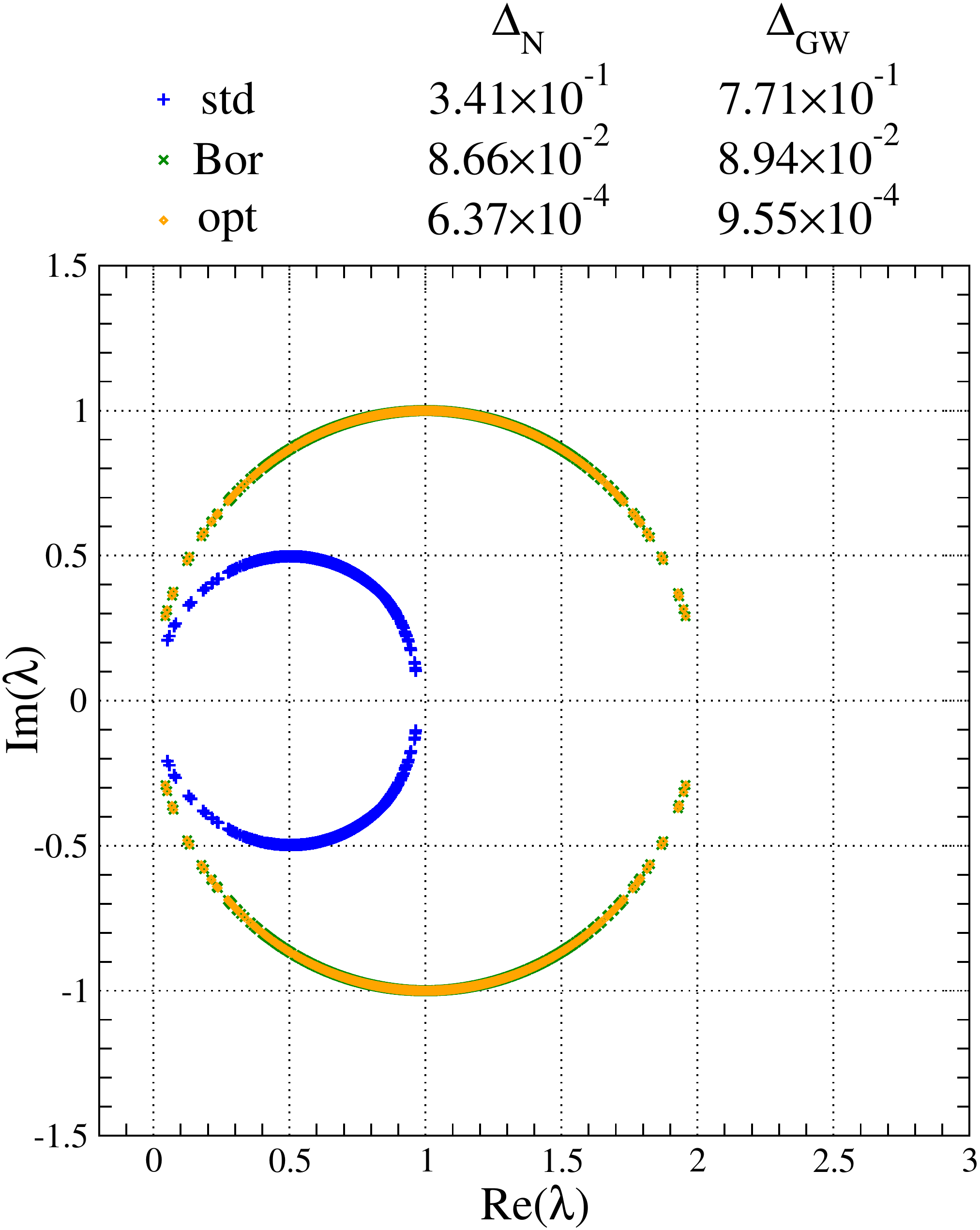}

}
\par\end{centering}
\caption{Spectrum of $\varrho D_{\mathsf{eff}}$ with a Wilson kernel
and a staggered Wilson kernel at $N_{s}=8$ in QCD\protect\textsubscript{4}
at $\beta=6$. \label{fig:deff-qcd}}
\end{figure}

Going beyond our study in Ref.\ \cite{Hoelbling:2016qfv}, we finally
discuss first results for staggered domain wall fermions in the setting
of quenched quantum chromodynamics in four dimensions. In Fig.\ \ref{fig:deff-qcd},
we show the eigenvalue spectrum of the
standard, Boriçi's and the optimal construction at $N_{s}=8$ on a
smeared $6^{4}$ configuration at $\beta=6$. The configuration was
smeared with one \textsc{ape} smearing iteration using $\alpha=0.65$.
For Wilson fermions we use a domain wall height of $M_{0}=1.4$, while
for staggered Wilson fermions $M_{0}=1$ remains the canonical choice.
As we are dealing with a topologically trivial configuration, we omit
$m_{\mathsf{eff}}$ in the figure labels.

Although it is too early for a full assessment of the chiral properties
of the effective operators without obtaining more statistics, we note
that on the present gauge configuration we observe a large reduction of
$\Delta_{\mathsf{N}}$ and $\Delta_{\mathsf{GW}}$ by typically more
than an order of magnitude for Boriçi's and the optimal construction
when using the staggered Wilson kernel. For the standard construction,
however, the Wilson-based construction shows better chiral properties.

\section{Conclusions}

The use of staggered domain wall fermions results in significantly
improved chiral properties in the Schwinger model, offering
the prospect of a computationally cheaper construction. In the setting
of four-dimensional quantum chromodynamics, our first results are
also very encouraging and warrant further investigations.

\paragraph*{Acknowledgments.}

C.~H.~is supported by DFG grant SFB/TRR-55. C.~Z.~is supported
by the Singapore International Graduate Award (SINGA) and Nanyang
Technological University.

\bibliographystyle{apsrev4-1}
\bibliography{literature}

\end{document}